# Experimental evidence for pressure-induced first order transition in cerium nitride from B1 to B10 structure type

Morten B. Nielsen,[a] Davide Ceresoli,[b] Jens-Erik Jørgensen,[c] Clemens Prescher,[d] Vitali B. Prakapenka,[d] Martin Bremholm[a]*

[a] *Center for Materials Crystallography, Department of Chemistry and iNANO, Aarhus University, Langelandsgade 140, 8000 Aarhus C, Denmark*

[b] *Center for Materials Crystallography and Institute of Molecular Science and Technology (CNR-ISTM), via Golgi 19, 20133 Milano, Italy*

[c] *Department of Chemistry, Aarhus University, Langelandsgade 140, 8000 Aarhus C, Denmark*

[d] *Center for Advanced Radiation Sources, The University of Chicago, 5640 S. Ellis Ave., Chicago, Illinois 60637, USA*

* Corresponding author. E-mail: bremholm@chem.au.dk

The crystal structure of CeN was investigated up to pressures of 82 GPa, using diamond anvil cell powder X-ray diffraction in two experiments with He and Si-oil as the pressure transmitting media. In contrast to previous reports, we do not observe the B2 (CsCl type) structure at high pressure. Instead, the structural phase transition, starting at 65 GPa, from the ambient rock salt B1 structure results in a distorted CsCl-like B10 structure, irrespective of the pressure medium. Our result unambiguously confirms two recent density functional theory (DFT) studies predicting the B10 phase to be stable at these pressures, rather than the B2 (CsCl type) phase previously reported. The B10 structure appears to approach the B2 structure as pressure is increased further, but DFT calculations indicate that an $L1_0$ structure (AuCu type) is energetically favored.

**Keywords:** rare earth nitride; phase transformation; high pressure; distorted structure; diamond anvil cell

## I. INTRODUCTION

The lanthanoid pnictides (LnPn) display a wide variety of interesting physical properties, such as the magnetocaloric effect,[1] metallic to semiconducting behavior co-existing with magnetic order,[2] making them interesting for applications in spintronics.[3] Very recently LaSb has also been shown to be a topological semimetal that displays extreme magnetoresistance.[4] The LnPn are also challenging to describe with density functional theory (DFT) and finding a correct way to handle the 4$f$ orbitals is the subject of many theoretical papers.[5,6] As such, this class of compounds is still the worthy subject of much active investigation and it is important that experiments that complement the theoretical studies are carried out.

At first glance, the crystallography of the LnPn series seems simple as they all crystallize in the $Fm\bar{3}m$ rock salt structure with Strukturbericht symbol B1, Fig. 1(a).[7,8a] However, upon application of high pressure, compounds in this family display rich and interesting structural chemistry, owing to the more complex orbitals of lanthanoid (Ln) ions. Instead of simply undergoing a pressure-induced phase transition from B1 to B2 ($P m\bar{3}m$ CsCl structure, Fig. 1(b)) like NaCl, distorted derivatives of B2 occur.[7] For many combinations using early Ln and heavy pnictogen (Pn) atoms, an $L1_0$ type structure ($P4/mmm$ symmetry, Fig. 1(c)) has been reported.[8-12] Here the Ln coordination is still eight-fold with equal bond lengths, but the Pn-Ln-Pn bond angles are not all equal, resulting in contraction in the *c*-direction. In some cases the $P4/mmm$ phase is followed by a further transition to the B2 structure.[11]

For the lightest pnictogen, nitrogen, it was long assumed that high pressure would just induce the B2 structure and many theoretical studies have predicted transition pressures for various lanthanoid nitrides (LnN).[13] This picture was first challenged in 2010 when Cynn *et al.* reported the results of an investigation of PrN up to 85 GPa where they observed a new distorted structure, the B10 type ($P4/nmm$ symmetry Fig. 1(d)).[14] In this structure, isotypic with high pressure BaO,[15] the coordination of the lanthanoid is still eight-fold, but with the Ln-ion shifted from the center of the pseudo-CsCl cell towards the *ab*-face, resulting in four short and four long bonds. The direction of the shift is reversed for the neighboring Ln. Shortly following the PrN result, Schneider *et al.* reported the same type of transition for LaN,[16] and the transition has since been rationalized in terms of a Peierls distortion.[17]

Shortly before the publication of the LaN result by Schneider *et al.*, a paper reporting experimental evidence for a (partial) B1 to B2 transition in CeN, using energy dispersive powder X-ray diffraction (ED-PXRD), was published by Staun Olsen *et al.*[18] This result was also backed up by theoretical calculations. The result seemed to fit with reports of the other Ce pnictides CeP and CeAs displaying a B1 to B2 transition at moderate pressures (note that no experimental data was shown in either report),[19] but also raised the question of why CeN displayed a direct B1 to B2 transition while its neighboring compounds LaN and PrN did not.

A possible resolution was suggested a year later in independent papers by Sahoo *et al.*[20] and Zhang *et al.*[21] Both groups investigated the phonon spectrum of B2 CeN at high pressure (60 and 80 GPa, respectively) and found negative frequencies indicating dynamic instability of this structure. By analyzing the real space displacements of Ce along the eigenvector with the negative frequencies, Zhang *et al.* showed how this led directly to the B10 structure and a lower energy.

In this paper, we report unambiguous experimental evidence for transformation of B1 CeN to the B10 structure, using angular dispersive powder X-ray diffraction (PXRD). The B10 *c/a'* tetragonal splitting is much smaller than that observed in for instance LaSb and CeSb,[10] which may be the reason it has been overlooked previously. Results from experiments using two very different pressure media, He and silicone oil (Si-oil), are in good agreement when lattice strain induced by non-hydrostatic pressure is included in the refinement models. The onset of the transition is ~65 GPa and we characterize the further evolution of the tetragonal structure up to 82 GPa in both experiments. The results are also compared to our own DFT calculations that provide indications that the B2 phase will not be reached as pressure is increased further, a possibility that has so far been overlooked for the Ln nitrides.

## II. METHODS

### A. Synthesis

Cerium mononitride was synthesized by heating Ce metal chips at 900 °C in a flow of gaseous ammonia for 15 hours, followed by furnace cooling to room temperature. The obtained powder samples were transferred to an argon glove box to prevent contact with moisture. Here the samples were ground together with a small amount of fine-grained Cu pressure standard and pressed to 10-15 μm thick foils before subsequent transfer to the diamond anvil cell chambers. A very faint (1-2 wt%) oxide impurity signal could be detected in both samples, which we attribute to very slight surface oxidation of the samples.

### B. High pressure diamond anvil cell experiments

*In situ* angular dispersive powder X-ray diffraction (PXRD) as a function of pressure was carried out in axial mode (transmission through the diamonds) at room temperature at beamline GSECARS 13-ID-D at the Advanced Photon Source (APS), using monochromatic X-rays with an energy of 40.000(2) keV, corresponding to a wavelength of $\lambda = 0.3100(2)$ Å and a MarCCD detector. The beam size on the sample was about 3x5 μm$^2$. Samples were loaded in Mao-Bell type symmetric diamond anvil cells (DACs) with 200 μm culets. Rhenium gaskets with a thickness of 250 μm were pre-indented to about 35 μm thickness prior to laser-cutting of a hole with a diameter of 90-100 μm into which the samples were loaded. The pressure transmitting medium (PTM) was either silicone oil (Si-oil) or helium; the latter loaded using the COMPRES/GSECARS gas loading system[22] at the APS. The use of both PTMs facilitates direct comparison with Staun Olsen *et al.*[18] and allowed investigation of the effect of PTM on the phase transition. Cu mixed into the sample was used as the internal pressure standard. The pressure was calculated from the refined unit cell parameter, using the equation of state of Dewaele *et al.*[23]

### C. Calibration and refinements

The experimental setup was calibrated using a LaB$_6$ standard (NIST SRM 660b). The 2D powder diffraction data was integrated in 4 degree azimuthal sections (90 sections in total) to enable refinement of anisotropic macrostrain. All refinements were carried out using the MAUD software,[24] starting from the guidelines described in Lutterotti *et al.*[25] and Wenk *et al.*[26] The "Radial diffraction in the DAC" model for lattice strain under non-hydrostatic conditions was then implemented. Despite its name, the model does not actually require any specific geometry (such as radial DAC diffraction) to be used, with the proviso that the correct coordinate system is specified in MAUD.[26,27] In our case, with axial DAC diffraction geometry, the correct coordinates are obtained by setting the "Omega" angle to 90 degrees instead of 0 degrees as done in Wenk *et al.*[26]

In the B1 region, intensities were modelled using arbitrary texture, the 2D equivalent to a Le Bail refinement, to get the most precise lattice parameters. After the onset of the phase transition, full Rietveld refinements were carried out. Peak broadening was modelled by refining one isotropic size term and anisotropic microstrain under the "Anisotropic no rules" model, using up to three terms if required to get a good peak description. No information from peak broadening is sought in this study. Structure drawings were made using the VESTA program.[28]

## D. Equation of state fitting

The pressure-volume (*PV*) behavior of the different CeN phases was fitted using the Birch-Murnaghan (BM) equation of state[29] (EoS), equation (1):

$$P = 3K_0 f_E (1 + 2f_E)^{5/2} \left\{ 1 + \frac{3}{2}(K_0' - 4)f_E + \frac{3}{2}\left[K_0 K_0'' + (K_0' - 4)(K_0' - 3) + \frac{35}{9}\right] f_E^2 \right\} \quad (1)$$

Where $f_E$ is the Eulerian strain, equation (1a):

$$f_E = \left[\left(\frac{V_0}{V}\right)^{2/3} - 1\right]/2 \quad (1a)$$

and $K_0$, $K_0'$, $K_0''$ and $V_0$ denote bulk modulus, first and second derivatives of bulk modulus with respect to pressure and volume, all at zero pressure. The DFT results required fitting using the full fourth order EoS (BM4) described above in order to produce a good fit, while the experimental B1 results were adequately fitted using the third order truncation (BM3) where the term proportional to $f_E^2$ is zero. Least squares fitting of the EoS to the *PV* data and extraction of parameters and covariance matrices was done using the program EoSFit 7c.[30]

## E. Theoretical calculations

DFT total energy calculations were performed within the plane-wave pseudopotential method, with Quantum-Espresso.[31] We generated an ultra-soft pseudopotential[32] for Ce including semi-core 5*s*, and 5*p* orbitals, in addition to valence 6*s*, 6*p*, 5*d*, 4*f* and the unbound 5*f* orbitals. The plane-wave and density cut-off were set to 55 Ry and 550 Ry, respectively. We used the semi-local PBE functional.[33]

The projected density of states (pDOS) in B1 CeN shows the states at the Fermi level have a strong N 2*p* character while the Ce 4*f* orbitals are empty. Therefore, Ce 4*f* orbitals, contrary to what is reported in literature, are not good candidates for the inclusion of the on-site Hubbard U correction.

Instead, we apply a U of 6.2 eV on the N 2*p* orbitals and in the following further motivate our choice. First, the recent ACBN0[34] formulation of DFT+U showed that in many perovskites and oxides, it is necessary to apply a large (6-8 eV) value of U on oxygen 2*p* orbitals, rather than on the transition metal 3*d* orbitals. The reason is that the metal-oxygen (and the metal-nitrogen bond) has too much covalent character in semi-local DFT. The Hubbard U term on O and N has the effect of cancelling the self-interaction of the electrons forming the metal-(O/N) bond, increasing its partial ionic character. This was proved in ref. 34 to lead to much improved structural and spectroscopic properties.

Unfortunately, the ACBN0 method is implemented only for norm-conserving pseudopotentials, whereas for ultra-soft pseudopotentials it lacks the augmentation charge, and underestimated the Hubbard U values. Nevertheless, we computed the ACBN0 Hubbard U values with our ultra-soft pseudopotentials and found a vanishing $U(Ce_{4f})$ value and a $U(N_{2p})$ of 4.5 eV. We then increased the value of $U(N_{2p})$ until we got a good agreement between the electronic band structure of B1 CeN (see Supplementary Material (SM),[35] Fig. S1) and that obtained with the Quasi-particle Self-consistent GW (QSGW) method,[36] as reported by Kanchana *et al.*[37]

The effect of this Hubbard U correction is twofold: first, at zero pressure the equilibrium volume of B1 CeN is reduced from 32.253 Å$^3$ to 31.666 Å$^3$, in much better agreement with experiments. Note that a recent very accurate and state-of-the-art Full-Potential Linearized Augmented-Plane Wave (FPLAPW) calculation of CeN reported an equilibrium volume 32.369 Å$^3$ without +U correction.[6] Second, the N 2$p$ states are pushed down in energy by ~2 eV (see SM,[35] Fig. S1), making the Ce-N bond a bit more ionic with respect to pure PBE. The mechanical stability of the B1, B2 and B10 phases was confirmed by phonon calculations using the frozen phonon method.

### III. RESULTS AND DISCUSSION

### A. Phase transitions

Selected PXRD data sets for CeN in Si-oil and He as a function of pressure are shown in Fig. 2 For clarity, only low angles are presented, see the SM[35] Fig. S2 for full angular range plots of both samples. As pressure is increased, the B1 and Cu peaks shift to higher angle as the lattice spacings decrease. At about 65 GPa, a clear phase transition can be seen in both data sets, proceeding slowly as pressure is increased further. At the highest pressure of 82 GPa, both samples are nearly phase pure B10 CeN, as shown by the diffraction peaks of the B10 reflections indicated by vertical tick marks in the bottom of each figure panel.

### B. B2 versus B10 structure

As stated in the introduction, Staun Olsen *et al.* used ED-PXRD to conclude that CeN undergoes a transition from the B1 to the B2 structure, starting at roughly 65 GPa.[18] This experimental result was then questioned in theoretical studies by Sahoo *et al.*[20] and Zhang *et al.*,[21] both of which used calculated phonon spectra to show that B2 CeN is unstable with respect to the B10 structure at pressures up to 80 GPa. A Rietveld refinement of the B10 model against angular dispersive PXRD data on CeN in Si-oil at 81.6(8) GPa is shown in Fig. 3, demonstrating that the B10 structure yields a satisfactory fit. In contrast, a B2 model cannot account for the reflections at 12.1 and 13.6 degrees 2θ and gets nearly all intensities wrong (see SM,[35] Fig. S3), resulting in the large residual and a $\chi^2$ which is an order of magnitude larger in the B2 model. This picture is repeated in the experiment on CeN in He (compare the bottom datasets in Fig. 2(a) and 2(b)). We therefore confirm that the B10 structural model is the correct one after the phase transition.

### C. *PV* refinement results

The volume per formula unit of CeN in the B1 and B10 structures as a function of pressure is shown in Fig. 4 alongside the results of the DFT calculations. The results for the two different pressure media are essentially identical within the errors for the B1 phase, while the B10 volume differs slightly between the two pressure media, which we attribute to differences in the uniaxial macrostrain present in the two samples at the large pressures. As expected for a calculation that uses the PBE functional, DFT overestimates the volume by 1-2 %. The volume drop at the phase transition is quite large at about 11% for both samples. A full list of obtained unit cell parameters versus pressure is given in the SM,[35] Tables ST1 and ST2.

As reported by Staun Olsen *et al.*,[18] the phase transition proceeds quite slowly and takes about 15 GPa to (almost) reach completion as can be seen in Fig. 5 where the refined phase fractions of B1 and B10 versus pressure for the two samples are shown. In the experiment using Si-oil as the PTM, the onset of the transition is at 64 GPa, while in the He PTM experiment the onset is at 67 GPa. Unit cell values for B10 at pressures below 69 GPa were extracted using Le Bail refinement as full Rietveld refinements before this pressure showed too large correlations between lattice parameters, phase fractions and Ce z-position.

### D. Anisotropic macrostrain in B1 CeN

While the integrated 1D data in Fig. 2(a) and 2(b) appear at first glance to correspond to the ideal cubic B1 CeN structure, strong evidence of non-hydrostatic lattice strain is evident when this model is Rietveld refined against the data, as the refinement of B1 CeN in Si-oil at 51.1(5) GPa shows in the top panel of Fig. 6(a). Here the (111) reflection at 6.5° needs to be shifted to lower angles (larger unit cell), while the (200) reflection at 7.6° needs a shift to higher angle (smaller unit cell). This very strong disagreement between B1 model and data cannot be resolved without inclusion of non-hydrostatic lattice strain in the model.

A significant improvement of the refinement quality is achieved by using the lattice strain model developed by Singh *et al.*,[27] as is evident in the bottom panel of Fig. 6(a). The implementation of this model in MAUD uses two refinable angles, $\alpha$ and $\beta$, to describe the angle between the direct beam and the primary strain direction and a refinable strain magnitude, $|Q_{hkl}|$, for each reflection.[26] In our case, $\beta$ is the angle between our direct beam and the primary stress direction while $\alpha$ is a counter-clockwise rotation of this vector around the beam.

One may consider two scenarios in which the direct beam and the primary stress direction are either coincident or perpendicular. In the former, the Debye-Scherrer rings will be uniformly shifted with *hkl* dependent magnitudes and $\beta$ should refine to zero, which is the case in the Si-oil experiment. In the latter, a different *d*-spacing will be observed as a function of azimuthal angle along the Debye-Scherrer ring. In this scenario, $\beta$ should refine to a value close to 90° and this is what we observe in the He experiment.

The azimuthal variation of *d*-spacing is often referred to as "wiggly" lines and is observed more often in radial DAC geometry, since the primary stress direction is most often along the diamond axis. This does not preclude its observation in the axial geometry though,[38] as long as the direction is close to parallel to the diamond faces. In general, it will depend on the specific position and size of the sample compared to the pressure chamber, the PTM and the actual pressure. It therefore varies between sample loadings (also when using the same PTM[38]) and is likely to change with pressure as the chamber shrinks. A representative 2D plot of intensity *vs* azimuthal angle (cake plot) showing "wiggly" lines for CeN in He at 61 GPa is presented in the SM Fig. S4, while an Si-oil data set at 63 GPa with straight lines is shown in the SM Fig. S5.[35]

Since the value of $|Q_{hkl}|$ depends on the *hkl* index, the reflections are shifted by unequal amounts, as if the data violates the cubic symmetry. The net result is usually a systematic overestimation of the volume of the unit cell,[39] and modelling of the effect is therefore important in order to obtain a correct equation of state.

In Fig. 6(b), bottom, the largest $Q_{hkl}$ values are plotted as a function of pressure for the two samples. From these values it is possible to estimate the size of the uniaxial component of the stress *t* using equation (2) according to Singh *et al.*[27]

$$t = 6G\langle Q_{hkl}\rangle f(x) \qquad (2)$$

Where G is the aggregate shear modulus and *f(x)* is a function of the elastic constants, usually approximated as one.[40] In the Reuss limit *G* is obtained from equation (3) as:

$$G = \frac{15}{6C'+9/C} \qquad (3)$$

With $C' = (C_{11} - C_{12})/2$ and $C = C_{44}$. We have obtained elastic constants by a linear interpolation of the elastic constants calculated by Sahoo *et al.* in the pressure range 0-75 GPa.[20] The absolute values change slightly if the elastic constants reported by Zhang *et al*. are used, and the *t* value reported here should therefore be seen as an estimated minimum value.

The resulting uniaxial *t* versus pressure is plotted in Fig. 6(b), top. This plot underlines why He is usually preferred over Si-oil as a PTM. Appearance of non-hydrostatic stress in He at slightly more than 30 GPa has also been observed by Takemura and Dewaele, using the same analysis method.[38] Strain is released at the onset of the phase transition as a result of the large volume drop, and its inclusion in the refinement is only significant until the phase fraction of B1 goes below 50%. The deformation of B1 CeN is greatest along the <111> direction, followed by <220> and <311> while the <200> deformation consistently refined within 1-2σ of zero and was therefore not included. The change of slope in the pressure evolution of *Q* for the Si-oil experiment at ~20 GPa seems to correlate with *t* reaching ~1.65 GPa and is likely caused by the onset of plastic flow in the sample.

As shown, the choice of PTM makes a large difference in terms of the observed macrostrain. However, the small difference in onset pressure between the two experiments is of the same magnitude as the difference in uniaxial stress. From this we conclude that macrostrain does not affect the transition pressure for this system and the transition mechanism is therefore unlikely to be strain driven.

### E. Pressure evolution in B10 CeN

The evolution of the unit cell axes of the B10 phase of CeN in the two PTMs versus pressure is shown in Fig. 7(a) alongside the DFT result. The volumes from the two experiments are within the uncertainties as the phase transition reaches completion. The DFT prediction systematically underestimates the *a* axis by 0.3%, while the *c* axis is overestimated by about 0.5%. The cause of the deviation of the axis trends at the lowest pressures in the Si-oil experiment is not clear to us.

As pressure is increased, the *c/a'* ratio of the pseudo-CsCl cell (highlighted in Fig. 1(d)) decreases slowly. This is plotted in Fig. 7(b), where the results from the two PTMs are again identical within error as the transition reaches completion. A simple linear interpolation of the experimental trend between 70 and 83 GPa yields an estimated crossing of *c/a'* = 1 at 147(10) GPa. DFT overestimates the *c/a'* ratio in the experimental range and does not reach *c/a'* = 1 even at 200 GPa. Instead, the DFT based *c/a'* ratio reaches a minimum at 200 GPa after which it begins to slightly increase again, see Fig. 7(b) inset.

To check that this non-unity *c/a'* ratio is not an artefact of DFT, we compared the energy of two cells where we forced the ratio to be 1.0017 and 1.0057, and found that the latter is lower in energy by about 100 meV at 240 GPa. This energy difference is small, but not negligible at room temperature and it indicates that the slight distortion from cubic is a true feature of the system.

The Ce z-coordinate extracted from the Rietveld refinements of the two samples versus pressure is presented in Fig. 8(a) alongside the DFT prediction. Here the two pressure media again yield different trends at lower pressures and we attribute this to fairly strong correlation in the He experiment with both the B1 phase and the B10 unit cell axes while the phase fraction of B10 is still low. This picture is also clear if the two extracted Ce-N bond lengths in the B10 structure are compared as a function of pressure, plotted in Fig. 8(b). Here the results from the two PTMs still overlap within the propagated error. It is interesting here to note that the short bond distance actually increases with pressure and then seemingly saturates, leaving the long bond length to slowly approach the short one.

In the DFT calculations, the z-coordinate progresses linearly to $z = ½$, reaching this position at about 200 GPa. This behavior is in stark contrast with the $c/a'$ ratio which merely saturates at this pressure. These two observations are taken to indicate that the pressures in excess of 200 GPa will still not yield the B2 structure, as otherwise the energy minimization in the tetragonal B10 structure should continue reducing $c/a'$. Instead, once Ce reaches $z = ½$, the symmetry of the structure reduces to P4/*mmm* (L1$_0$ type, Pearson symbol *tP2*), observed at high pressure for a large number of lanthanoid pnictides, such as LaP, PrP and NdP,[8a] LaAs,[12] LaSb and CeSb,[10] and finally CeBi.[11] The structure has also been predicted theoretically for LaP, LaAs, LaSb and LaBi.[8b] As far as we are aware, this possibility has not been considered for any LnN system previously and certainly warrants further study.

### F. Equation of state results

The fitted BM3 EoS results for the combined experimental data (both PTMs) and the BM4 DFT B1 results are displayed in Fig. 9, showing the covariance ellipses in $K_0$ versus $K_0'$ and $V_0$ versus $K_0$ (inset). The center values are listed in Table 1 alongside the DFT B10 and B2 results. Reference values from other studies are also plotted in Fig. 9 and listed in Table I.

A BM4 fit of the B1 DFT results yields values of $V_0$ and $K_0$ that overlap with previous reports that used the BM3 EoS, but with a slightly larger $K_0'$. Our experimental results produce a $V_0$ that agrees with literature values but observe a substantially lower bulk modulus, $K_0$, and correspondingly larger $K_0'$ than those reported previously for CeN. The results for $K_0$ and $K_0'$ do however overlap with the experimental results reported for LaN.[16] If we fix $K_0' = 4$, we obtain $V_0$ and $K_0$ values that overlap with those of Staun Olsen et al.,[18] but an *f-F* plot of the experimental data (see SM,[35] Fig. S6) clearly shows this assumption to be invalid and that a refinable $K_0'$ is needed to describe the observations.

Fitting the BM4 EoS to the B10 DFT calculation results in a much lower ambient pressure bulk modulus than the B1 phase, but also a large $K_0'$, with the result that at the transition pressure (65 GPa), the bulk moduli of the two phases are virtually identical while the volume of the B10 phase is 12% lower. A similar trend was reported by Sahoo et al. for B10 CeN, though the absolute values are not in agreement.[20]

The total energies per formula unit as a function of volume for the B1, B10, L1$_0$ and B2 structures obtained from the DFT calculations are plotted as points in Fig. 10 with the fitted BM4 equations of state as full lines. As reported by Sahoo et al. and Zhang et al., the enthalpy at 0 K of the B10 structure becomes lower than that of the B1 structure as pressure is increased (Fig. 10 inset). The EoS fits predict a transition pressure of 53(1) GPa, somewhat lower than the experimentally observed onset, but completely consistent with the report by Sahoo *et al.*, while Zhang *et al.* predicted an even lower transition pressure of 37.5 GPa.

As the volume per formula unit decreases below 20 Å$^3$, the B10 and B2 total energies approach each other (here noting that above 200 GPa, B10 and L1$_0$ are the same). Whether their fitted enthalpies cross or not changes depending on the EoS order used to carry out the fitting (BM3 or BM4). When the calculated enthalpy of the B10, L1$_0$ and B2 phases are compared at pressures above 200 GPa (see SM,[35] Fig. S7), one immediately notes that the enthalpy differences between them are so small that changing the magnetic order is likely to lead to shifts in energy of the same magnitude. As an exhaustive exploration of the possible magnetic structures of the B10 and other phases is beyond the scope of this work, we believe that more accurate calculations using a more advanced hybrid potential DFT approach would be worth carrying out. This would provide the grounds for definite statements about the lowest enthalpy. Furthermore, as alluded to in section 3.E, the structural optimization of the B10 structure seems to indicate that another intermediate phase of P4/*mmm* symmetry could exist between the B10 and B2 phases. Simply performing an enthalpy comparison of B10 and B2 without searching for alternatives should be avoided, as they risk overlooking the correct structure.

## IV. CONCLUSION

We have presented clear experimental evidence that the correct high pressure structure of CeN is the B10 type. This brings CeN in line with the behavior observed in the neighboring compounds LaN and PrN and suggests a larger structural trend in the LnN series. We have undertaken experiments to investigate whether this transition continues in all the higher lanthanoids.

The B1 to B10 transition begins at ~65 GPa and is sluggish in both He and Si-oil, as expected of a first-order transition. DFT predicts a transition at 53 GPa. Results from both PTMs are consistent with each other, but only as long as the lattice strain due to non-hydrostatic stress is modelled. There can be many good practical reasons for using a poorly hydrostatic PTM in an experimental study. Gas loading systems may not be available, samples may be very air sensitive (the case for all LnPn compounds) and even the best PTMs, such as the noble gasses, eventually become non-hydrostatic if high enough pressures are reached. If experimental results are to be compared with DFT calculations that are inherently hydrostatic, the uniaxial component of the stress is therefore an important factor to take into account.

Upon further pressure increase, the B10 structure continuously changes towards B2, but DFT calculations predict that the *c/a'* ratio does not reach unity, even at 300 GPa, though the Ce *z*-coordinate saturates at ½ already at 200 GPa. This suggests that the true structure above 200 GPa is of the L1$_0$ type, observed in heavier LnPn, but not yet for the nitrides.

The $K_0$ and $K_0'$ obtained from a BM3 EoS fit to the experimental B1 CeN data overlap with those reported for LaN. Previous literature reports and our own DFT calculations predict a higher $K_0$ and a lower $K_0'$. The relative total enthalpies of the B10 and B2 phases are sensitive to the EoS formulations used and the two structure types continue to have near identical enthalpies even at very high pressures. DFT calculations using more expensive hybrid approaches that fully account for the possible magnetic structures of the possible phases are likely needed before a definitive statement about the true ground state structure can be made. At the same time, the possibility of intermediate alternative structures should also be kept in mind.

Finally, we note that the reported high pressure behavior of CeP and CeAs also differ from the systematic trends observed in the early Ln pnictides.[19] As these reports were based on laboratory energy dispersive data, it possible that the relatively small tetragonal splitting (see Fig. 3) could have been overlooked. It may therefore be worth revisiting these compounds using modern synchrotron based setups.

## SUPPLEMENTARY MATERIAL

See supplementary materials for the electronic band structure of B1 CeN at equilibrium volume, Full angle region of selected datasets from the two sample runs, Table of refined parameters versus pressure for the two experiments, Obs.-calc. diagram of a B2 model refined against experimental data at 82 GPa, Cake plot of diffraction data of CeN in He PTM at 61 GPa and CeN in Si-oil PTM at 63 GPa, *f-F* plots for the B1 CeN results, Relative enthalpies of the B10, B2 and L1$_0$ phases in the pressure range 140 to 300 GPa.

## ACKNOWLEDGEMENTS

The authors gratefully acknowledge beamtime at APS GSECARS. This work was performed at GeoSoilEnviroCARS (Sector 13), Advanced Photon Source (APS), Argonne National Laboratory. GeoSoilEnviroCARS is supported by the National Science Foundation - Earth Sciences (EAR-1128799) and Department of Energy- GeoSciences (DE-FG02-94ER14466). Use of the COMPRES-GSECARS gas loading system was supported by COMPRES under NSF Cooperative Agreement EAR 11-57758 and by GSECARS through NSF grant EAR-1128799 and DOE grant DE-FG02-94ER14466. This research used resources of the Advanced Photon Source, a U.S. Department of Energy (DOE) Office of Science User Facility operated for the DOE Office of Science by Argonne National Laboratory under Contract No. DE-AC02-06CH11357. The Danish Agency for Science, Technology and Innovation (DANSCATT) is acknowledged for supporting the synchrotron activities. MB thanks the Villum Foundation and CMC for funding. CMC is a Center of Excellence funded by the Danish National Research Foundation (DNRF93). Camelia Stan is thanked for fruitful discussions.

TABLE I

Values of the fitted BM3 EoS zero pressure B1 CeN volume $V_0$, bulk modulus $K_0$, its first and second pressure derivatives $K_0'$ and $K_0''$ (when a BM4 EoS was used) from this work and literature. Experimental values for LaN are also included for comparison.

| Reference | Source | Phase | $V_0$ (Å$^3$) | $K_0$ (GPa) | $K_0'$ | $K_0''$ (GPa$^{-1}$) |
|---|---|---|---|---|---|---|
| This work | Exp. | B1 | 31.647(8) | 135.2(12) | 5.35(8) | |
| Staun Olsen et al.[18] | Exp. | B1 | 31.65(15) | 156(3) | 4 (fixed) | |
| Schneider et al.[16] | Exp. | B1 LaN | 37.325 | 135.5(3.5) | 5.0(5) | |
| This work | DFT | B1 | 31.56(5) | 165(2) | 4.62(14) | -0.063(6) |
| | | B10 | 31.2(3) | 59(6) | 7.2(6) | -0.36(13) |
| | | B2 | 27.27(4) | 168.3(12) | 4.26(4) | -0.0288(12) |
| Kanchana et al.[37] | DFT | B1 | 31.627 | 158.1 | 3.3 | |
| Staun Olsen et al.[18] | DFT | B1 | 31.702 | 158.1 | 3.3 | |
| Sahoo et al.[20] | DFT | B1 | 31.85 | 168.2 | 4.02 | |
| | | B10 | 30.55 | 99.1 | 4.9 | |
| | | B2 | 28.39 | 159.3 | 4.47 | |
| Zhang et al.[21] | DFT | B1 | 31.608 | 163.0 | 3.96 | |
| Topsakal and Wentzcovitch[6] | DFT | B1 | 32.345 | 150 | Not stated | |

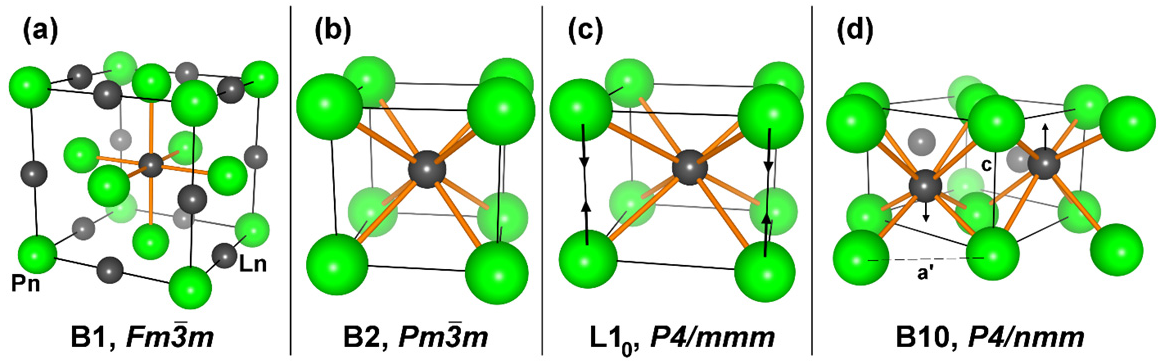

FIG 1. (Color online) (a) The B1 type rock salt structure displayed by all LnPn at ambient pressure. (b) The B2 type CsCl structure. (c) The L1$_0$ type structure displayed by many LnPn at high pressure, with the compressed *c*-axis highlighted. (d) The B10 type structure reported for LaN and PrN. Two neighboring pseudo-B2 cells are highlighted in front, while the *P4/nmm* unit cell is shown by the full lines.

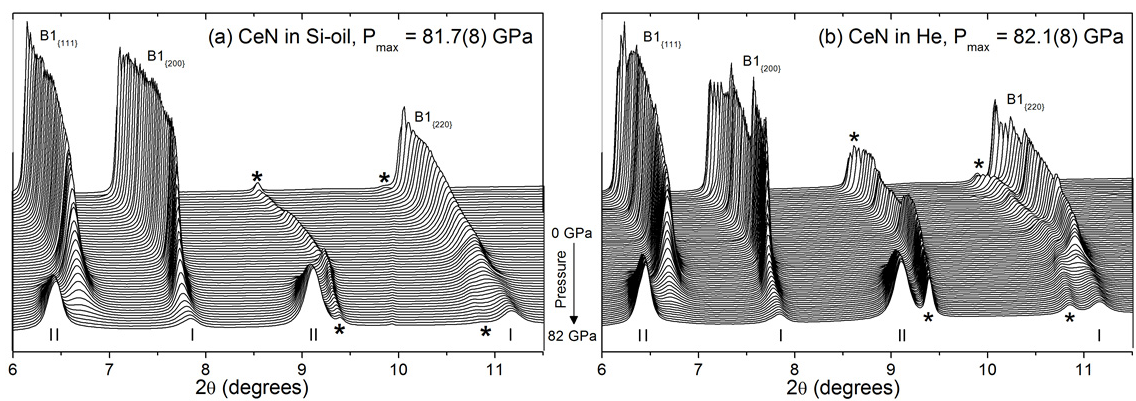

FIG. 2. The low angle region of selected datasets from the two sample runs with increasing pressure (spacing not to scale). The full refined region extends to at least 16° for both samples. (a) CeN in Si-oil, 53 out of 104 datasets shown. (b) CeN in He, all 67 datasets shown. Asterisk mark diffraction from the Cu pressure standard, tick marks indicate B10 peak positions in the bottom dataset.
17

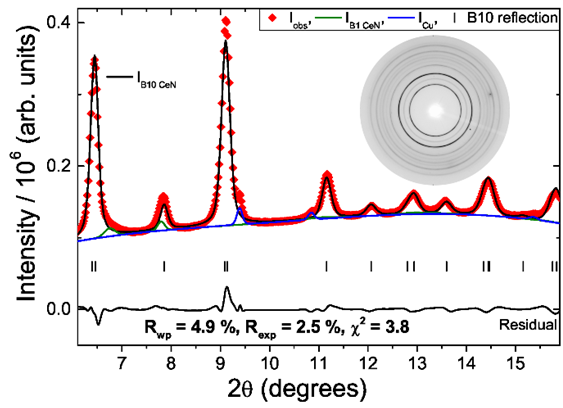

FIG. 3. (Color online) Rietveld refinement of CeN in Si-oil PTM at 81.6(8) GPa using the B10 model, with the 2D data as an inset. Several reflections that are forbidden in the B2 structure (at 12 and 13.5 degrees) are described by the B10 structure, and the total intensities of all reflections are in good agreement with B10 as well.

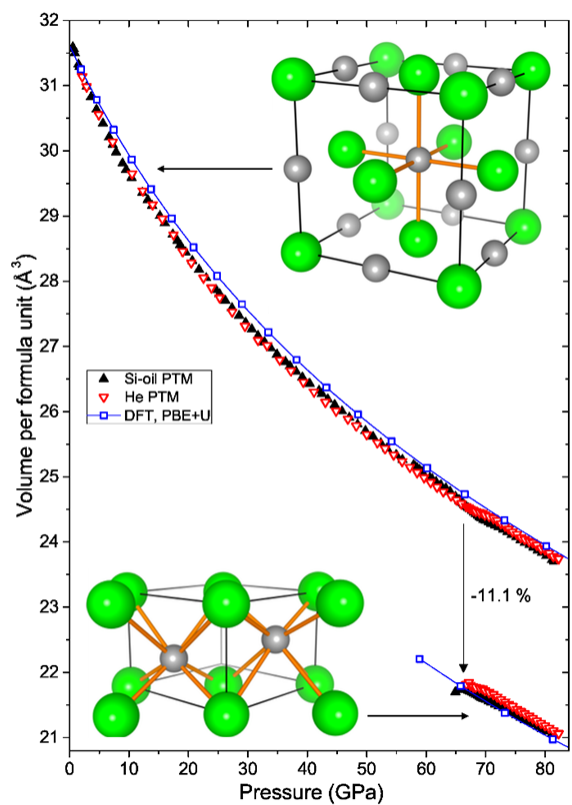

FIG. 4. (Color online) Volume per formula unit versus pressure for the two structures displayed and the two samples measured in different PTMs. Error bars are smaller than the symbols. At the phase transition the volume drop is about 11 % for both samples.

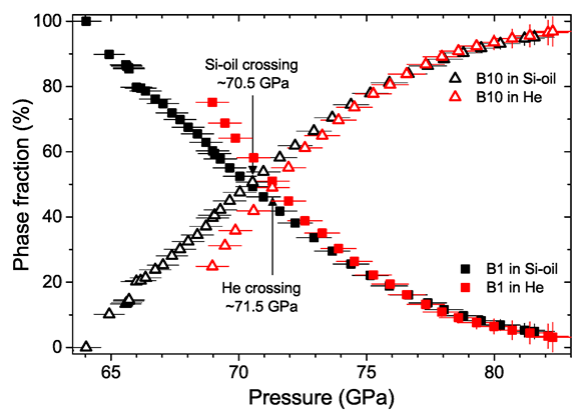

FIG. 5 (Color online) Phase fraction of the B1 and B10 structures in the two samples versus pressure. Rietveld refinements of the He PTM sample are unstable below 69 GPa.

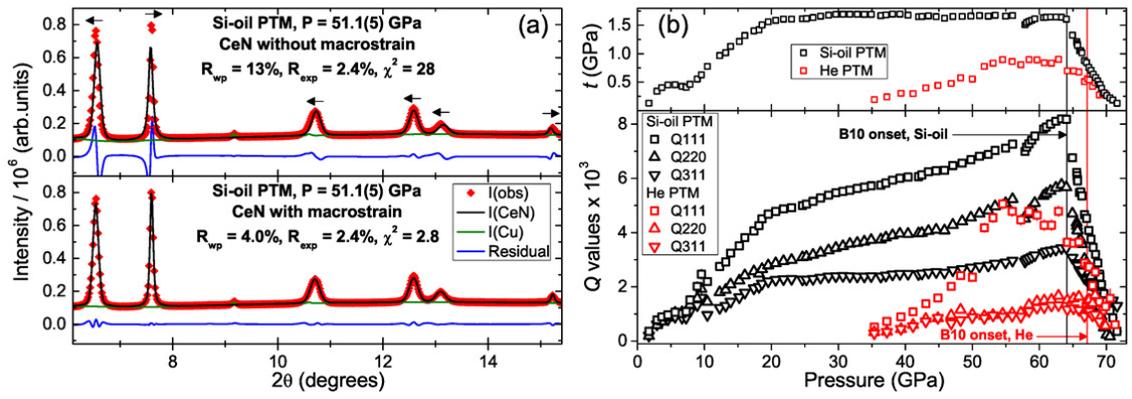

FIG. 6. (Color online) (a) Comparison of Rietveld refinements of CeN at 51.1(5) GPa, without (top) and with (bottom) modelling of anisotropic macrostrain. Arrows in the top panel highlight which direction the model reflections need to shift in order to improve the fit. No expanding or shrinking of the cubic cell can satisfy all requirements. (b) Bottom: Evolution of macrostrain parameters for different reflections as a function of pressure in the two samples. The vertical lines highlight the onset of the B1 to B10 transition in the different samples. Top: Derived estimate for the uniaxial stress $t$.

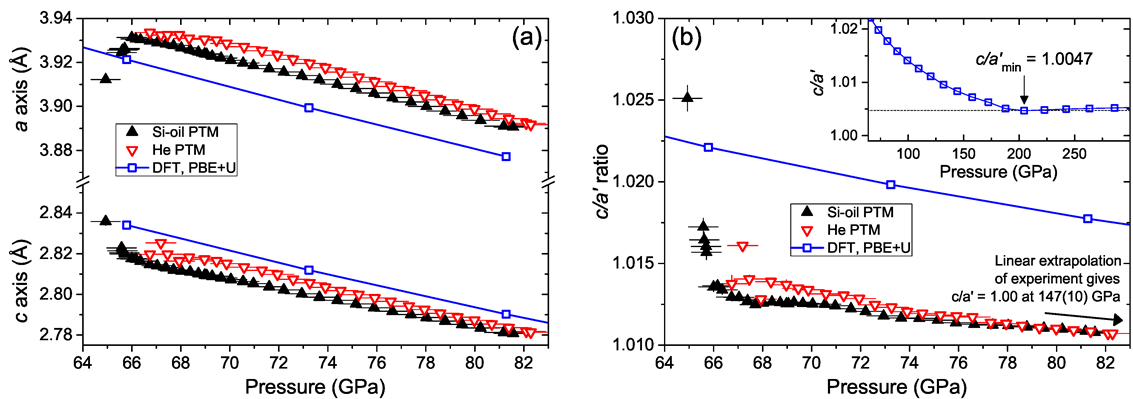

FIG. 7. (Color online) (a) B10 unit cell parameters versus pressure for the two experiments and from DFT. Vertical error bars are smaller than the symbols. (b) $c/a'$ ratio versus pressure for the two experiments and DFT. The reason for the different trend at the lowest pressures in the Si-oil experiment is unknown. Right inset: DFT results to the highest pressure calculated, showing saturation in the ratio above 200 GPa.

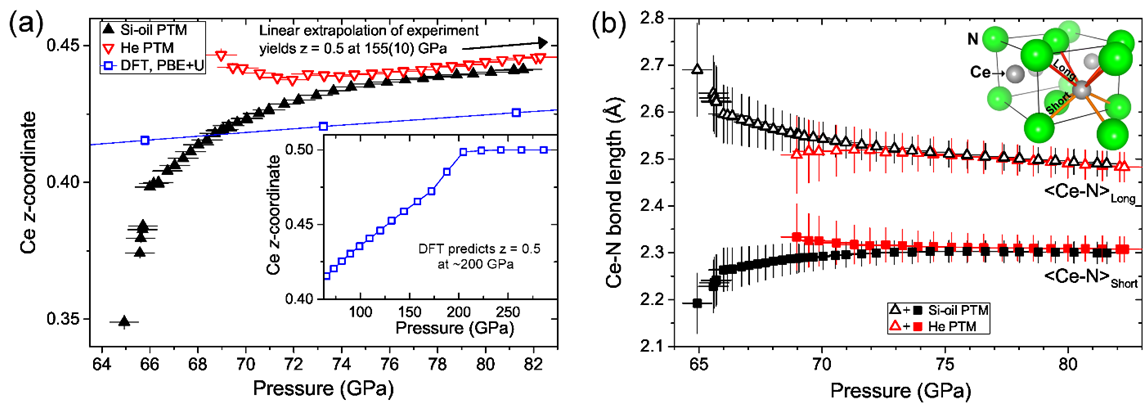

FIG. 8. (Color online) (a) Fractional z-coordinate of Ce in the B10 structure versus pressure in the two samples. Linear interpolation from 75 GPa yields a crossing of 0.5 at a pressure of 155(10) GPa. (b) Long and short Ce-N bond lengths versus pressure in the two samples.

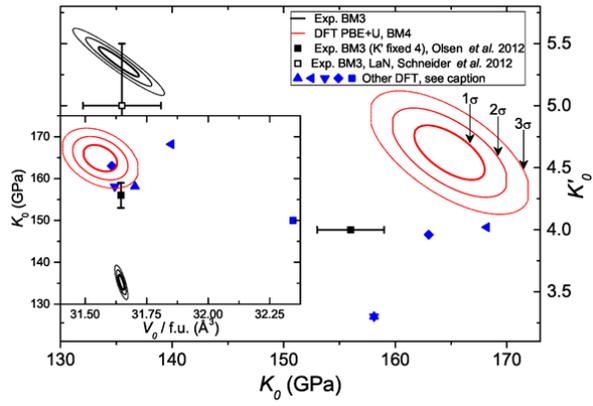

FIG. 9. (Color online) Bulk modulus, $K_0$, and first pressure derivative, $K_0'$, correlation plot with literature values for B1 CeN. DFT symbols, left to right, represent references 18, 20, 37, 21 and 6. Inset: Ambient pressure volume, $V_0$, and bulk modulus, $K_0$, correlation plot with literature values.

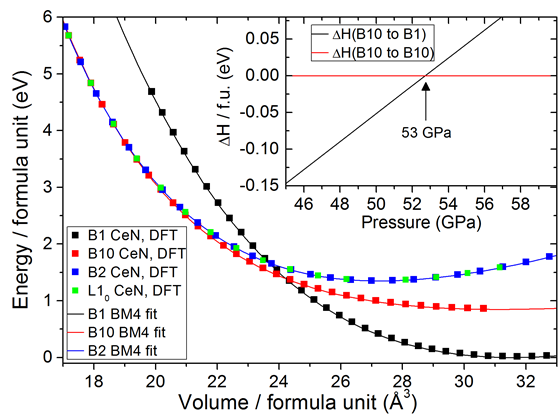

FIG. 10. (Color online) Total energies per formula unit versus volume from DFT for the phases B1, B10 and B2, normalized to the B1 zero pressure energy. Full lines represent the respective fitted BM4 EoS. Inset: Enthalpy difference from the B10 phase versus pressure, with the B10 structure having the lowest enthalpy above 53 GPa.

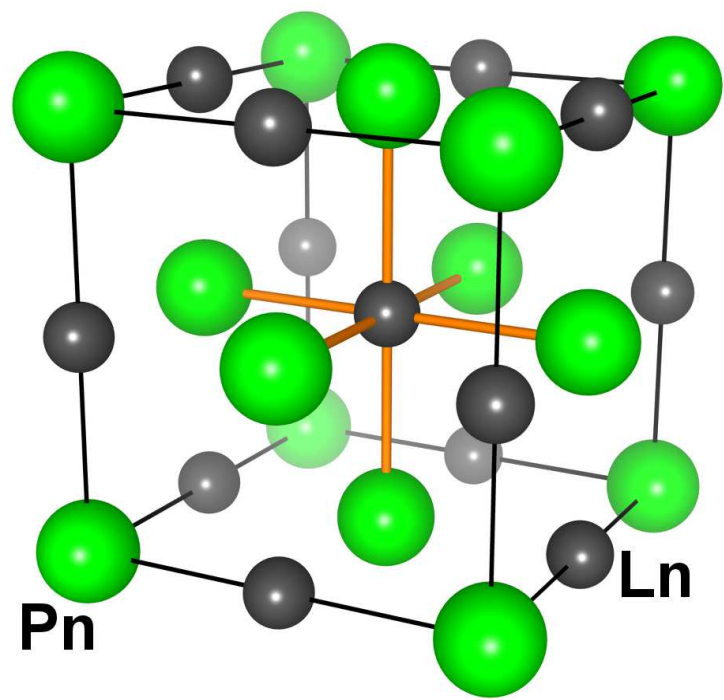 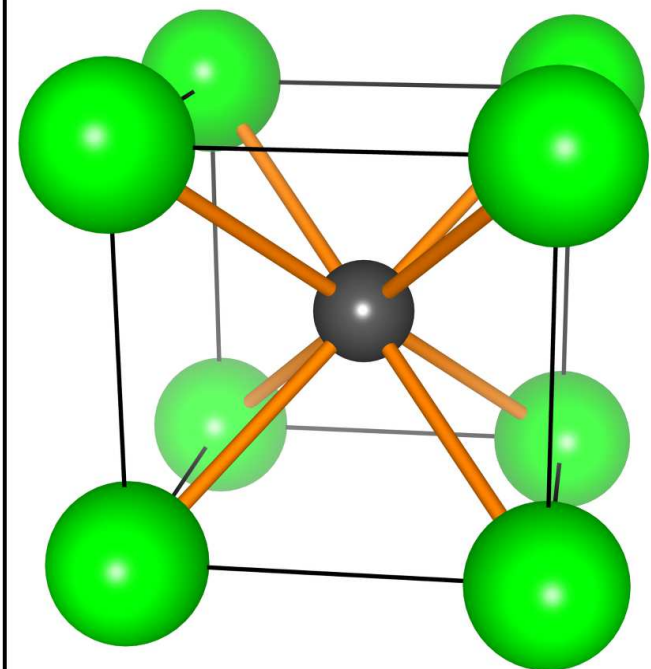 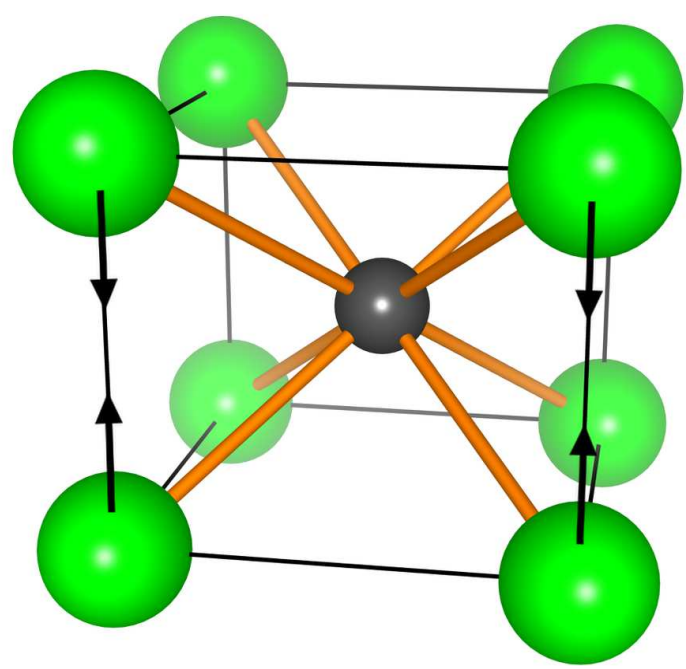 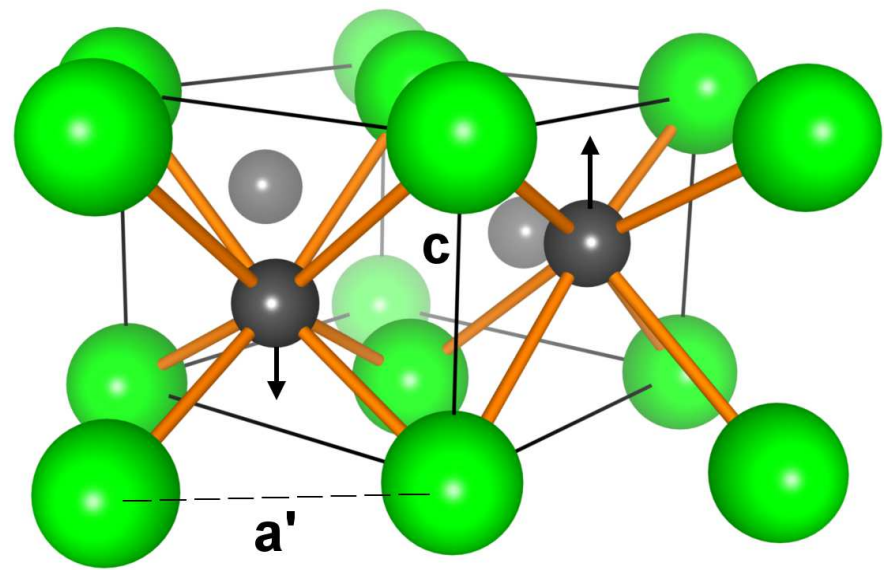

**(a)** B1, *Fm$\bar{3}$m*  **(b)** B2, *Pm$\bar{3}$m*  **(c)** L1$_0$, *P4/mmm*  **(d)** B10, *P4/nmm*

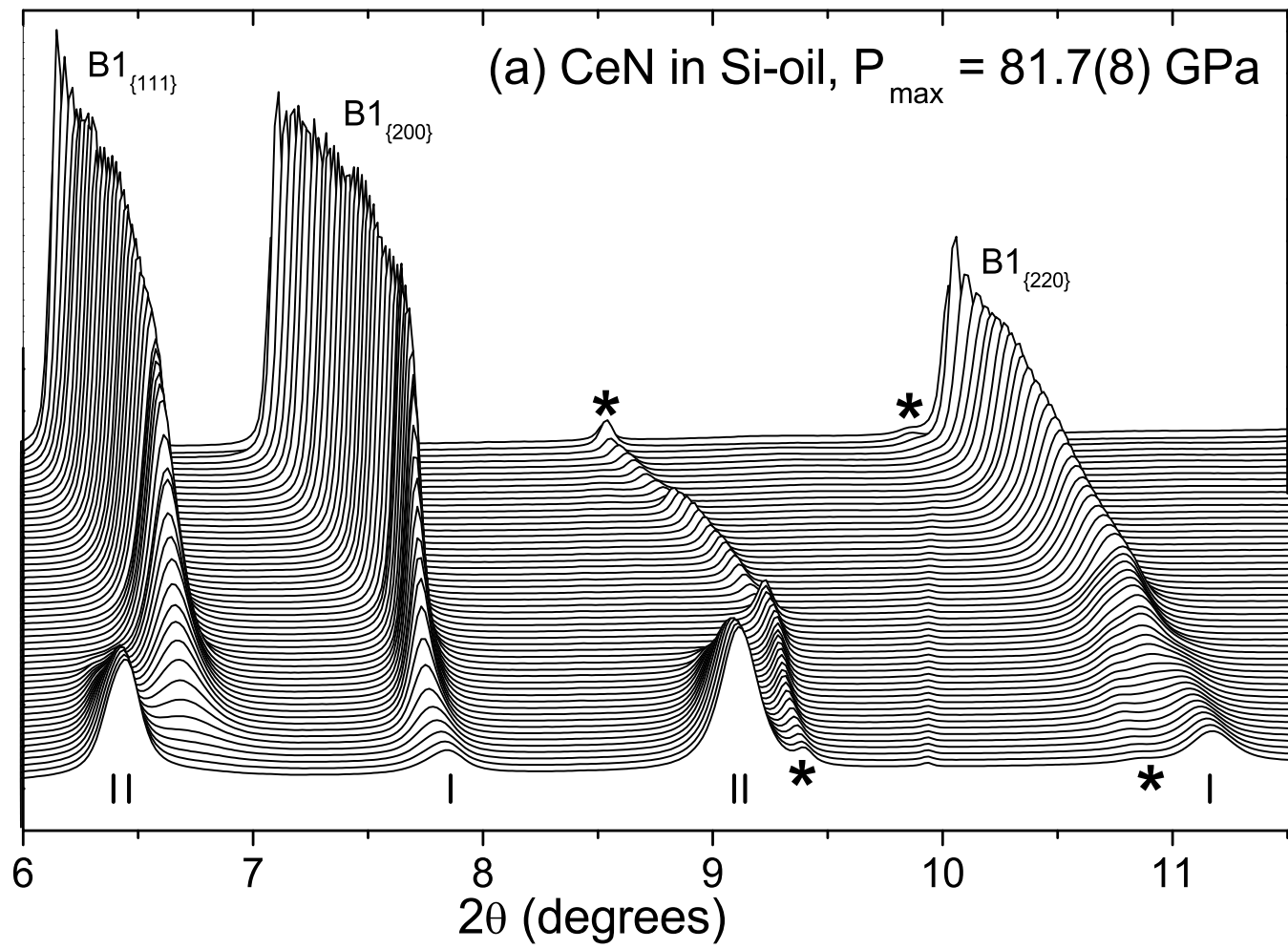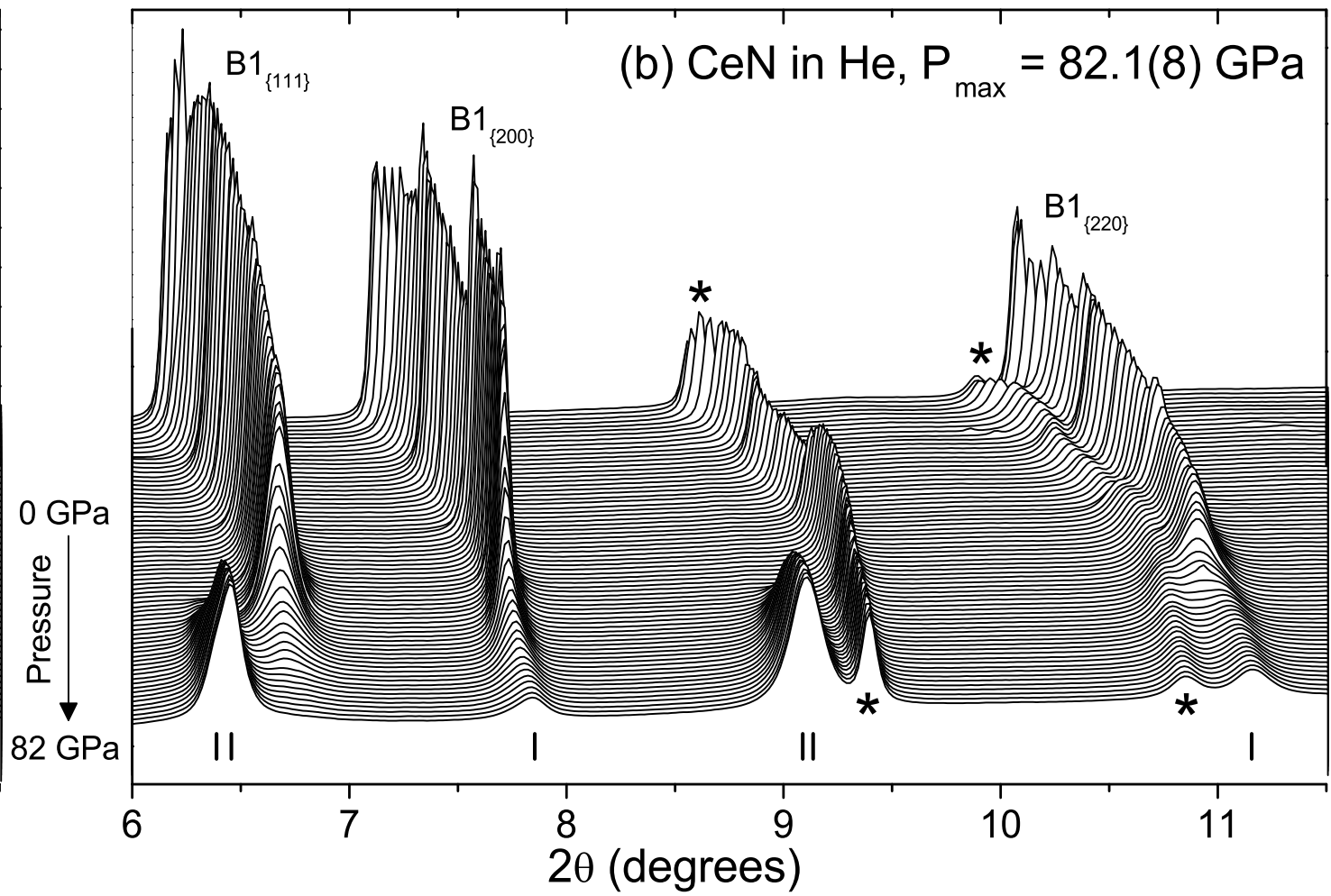

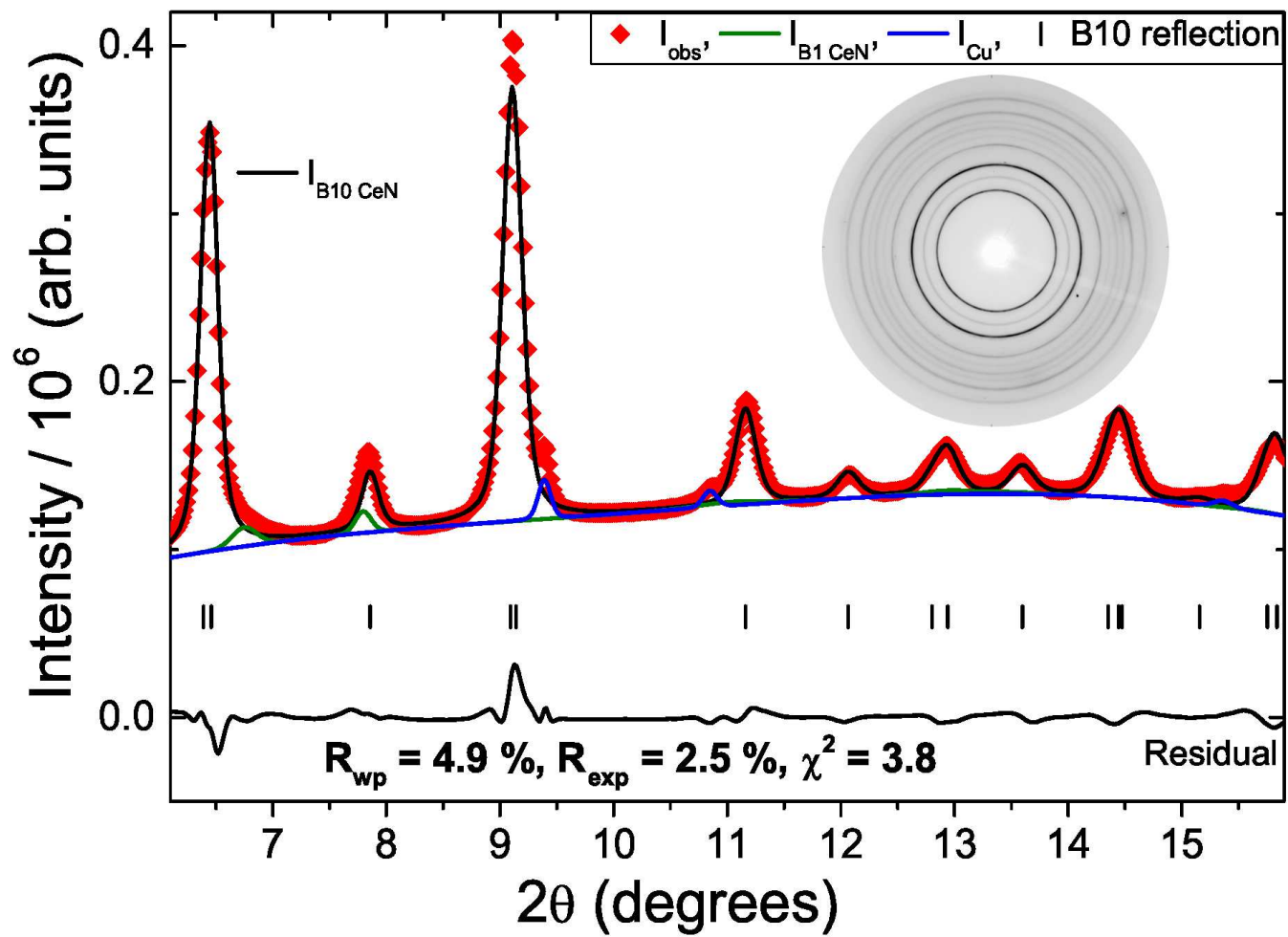

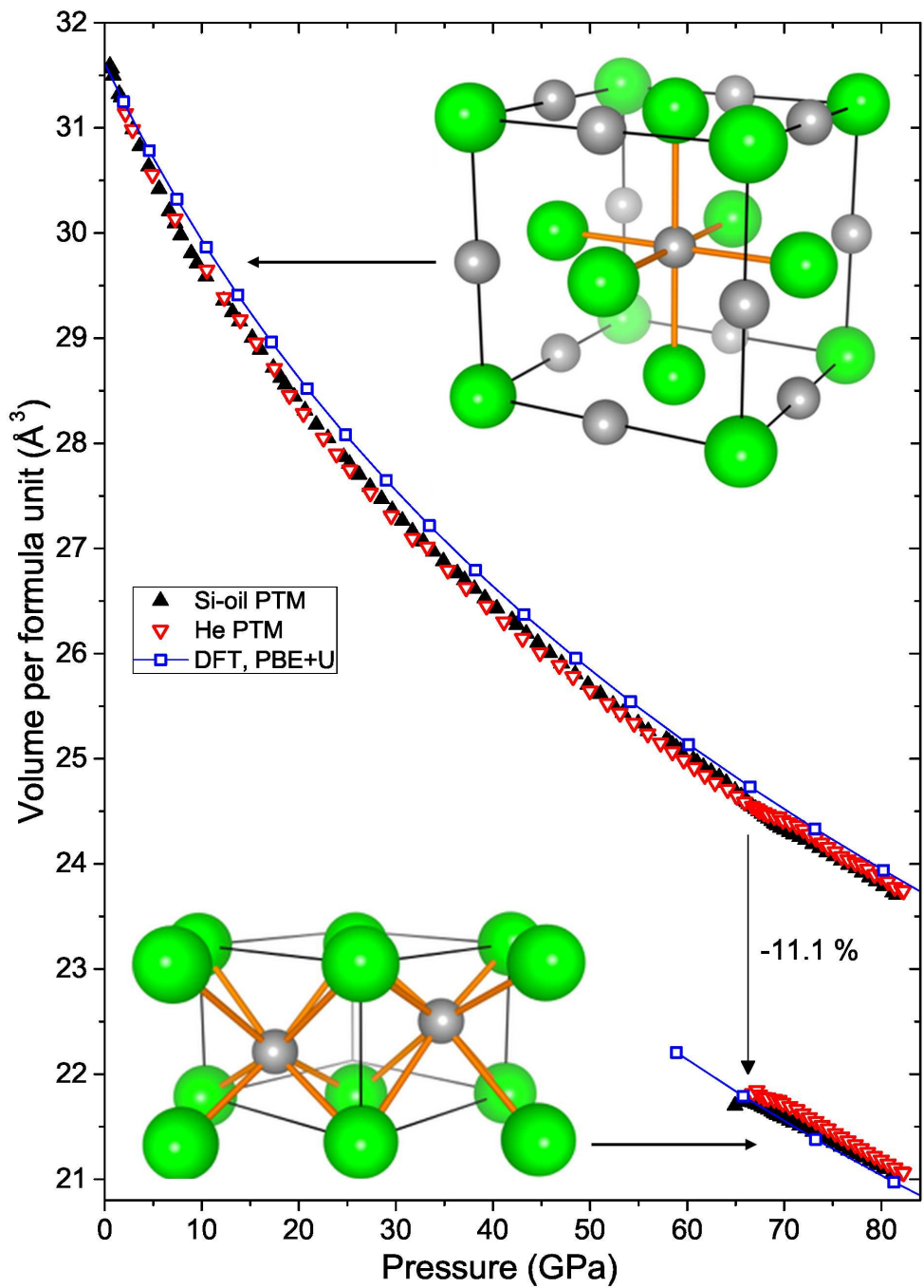

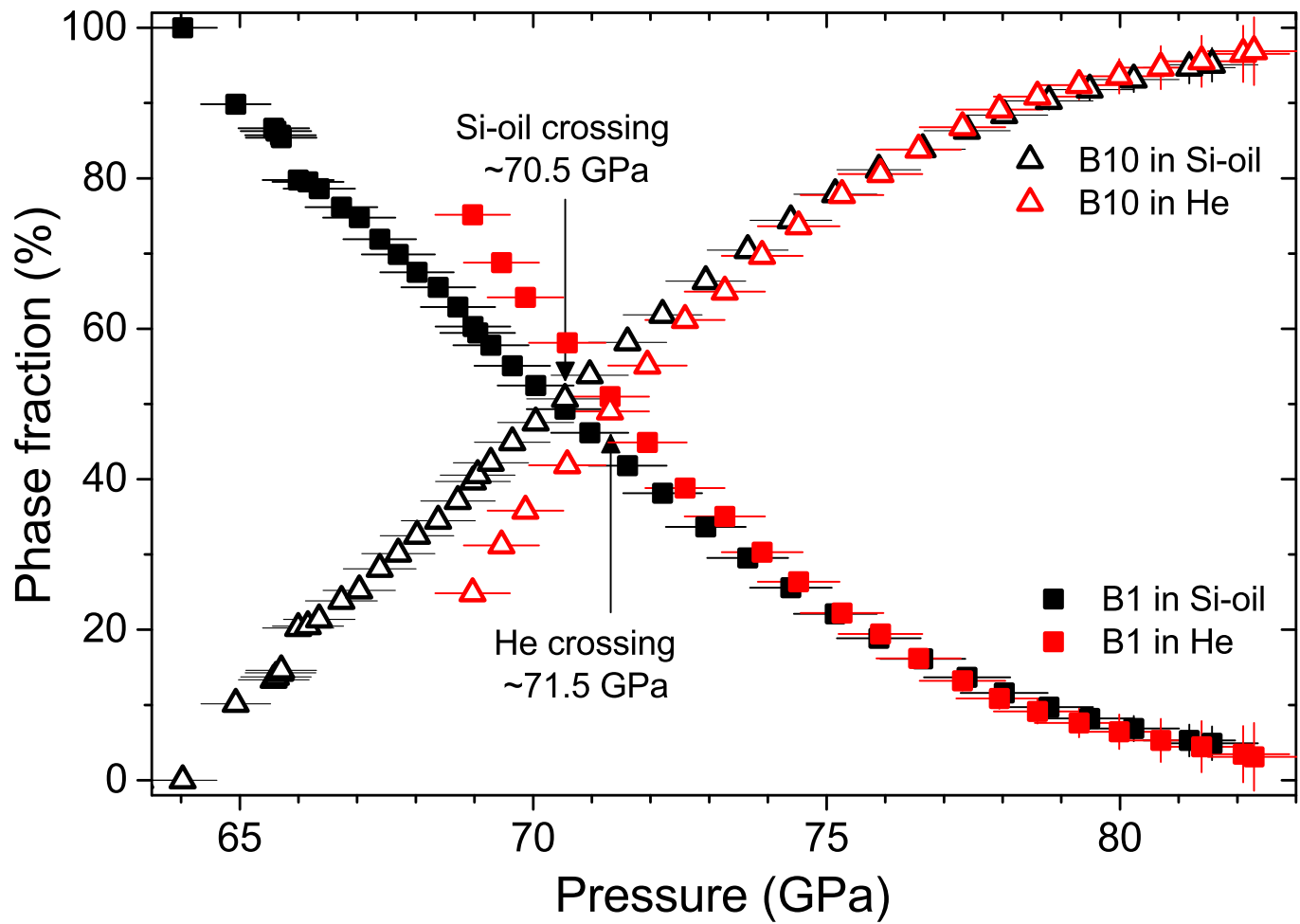

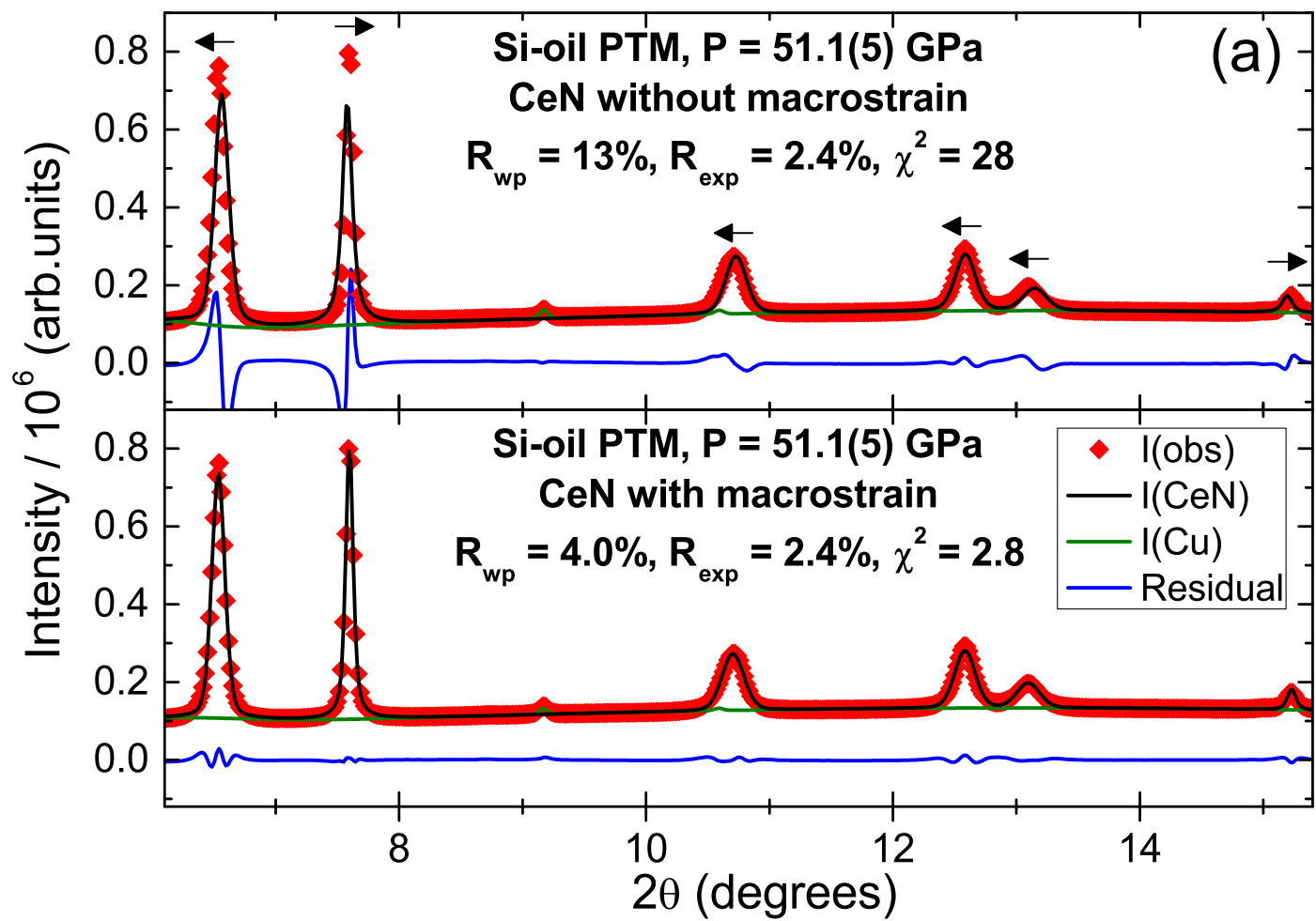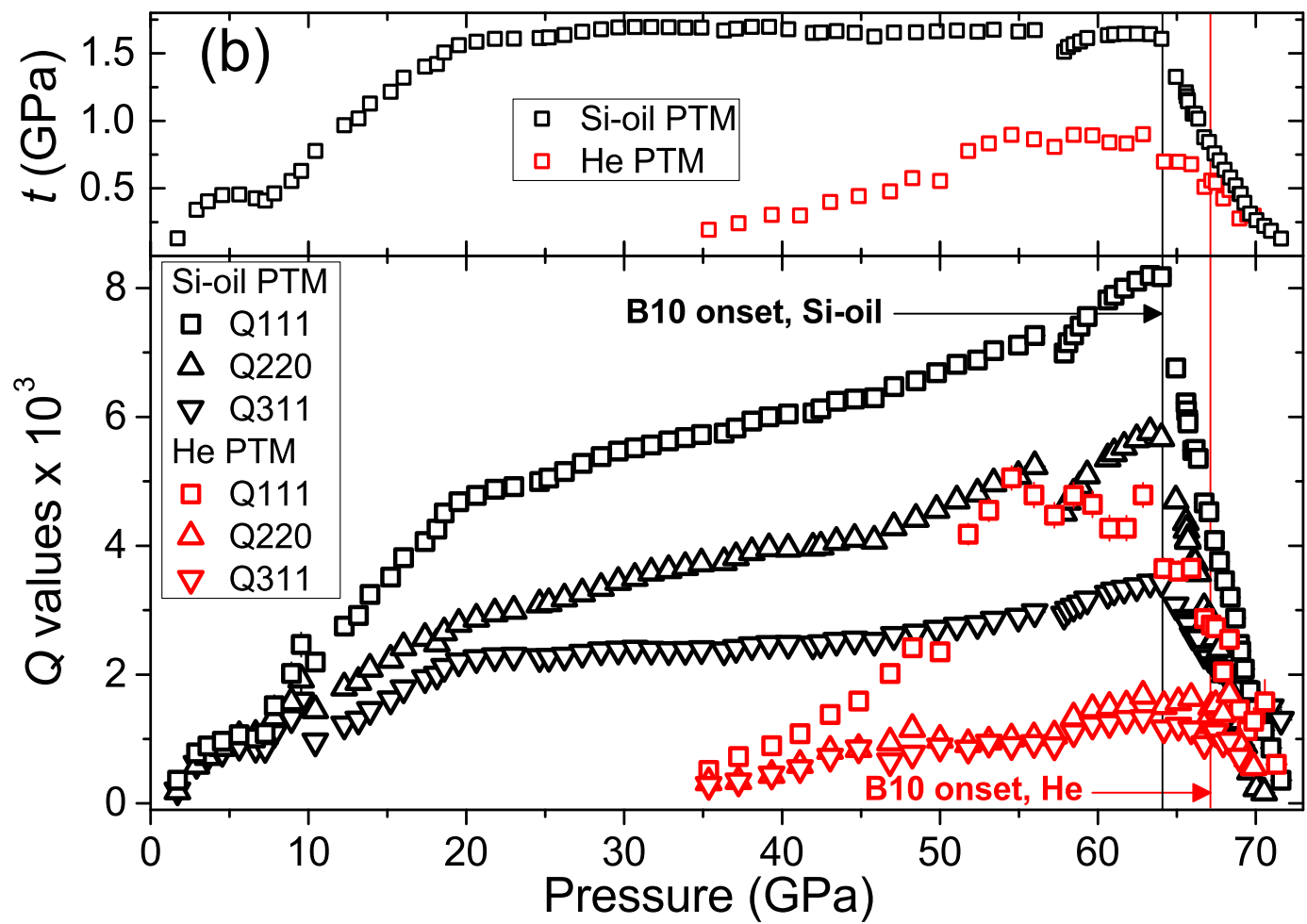

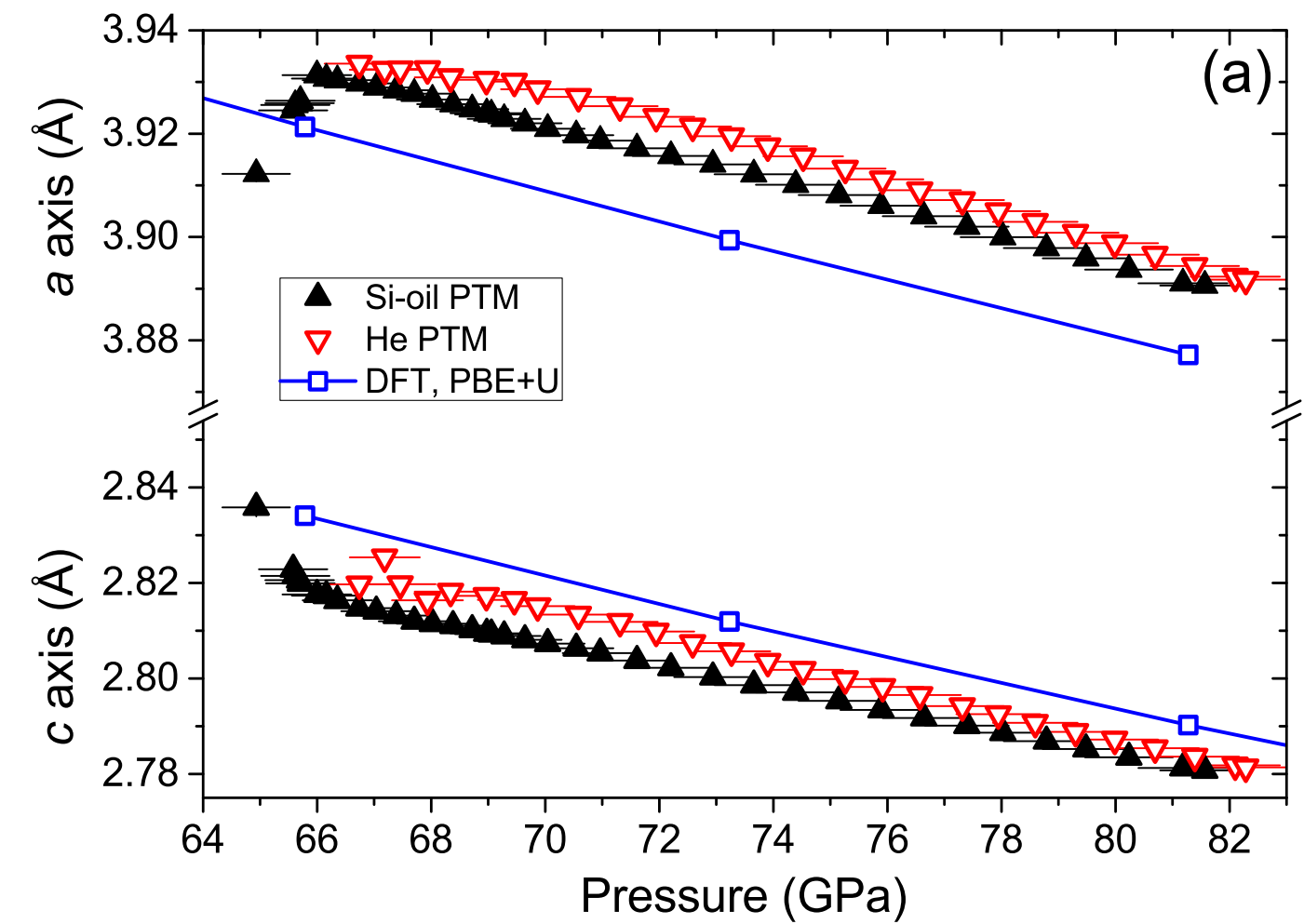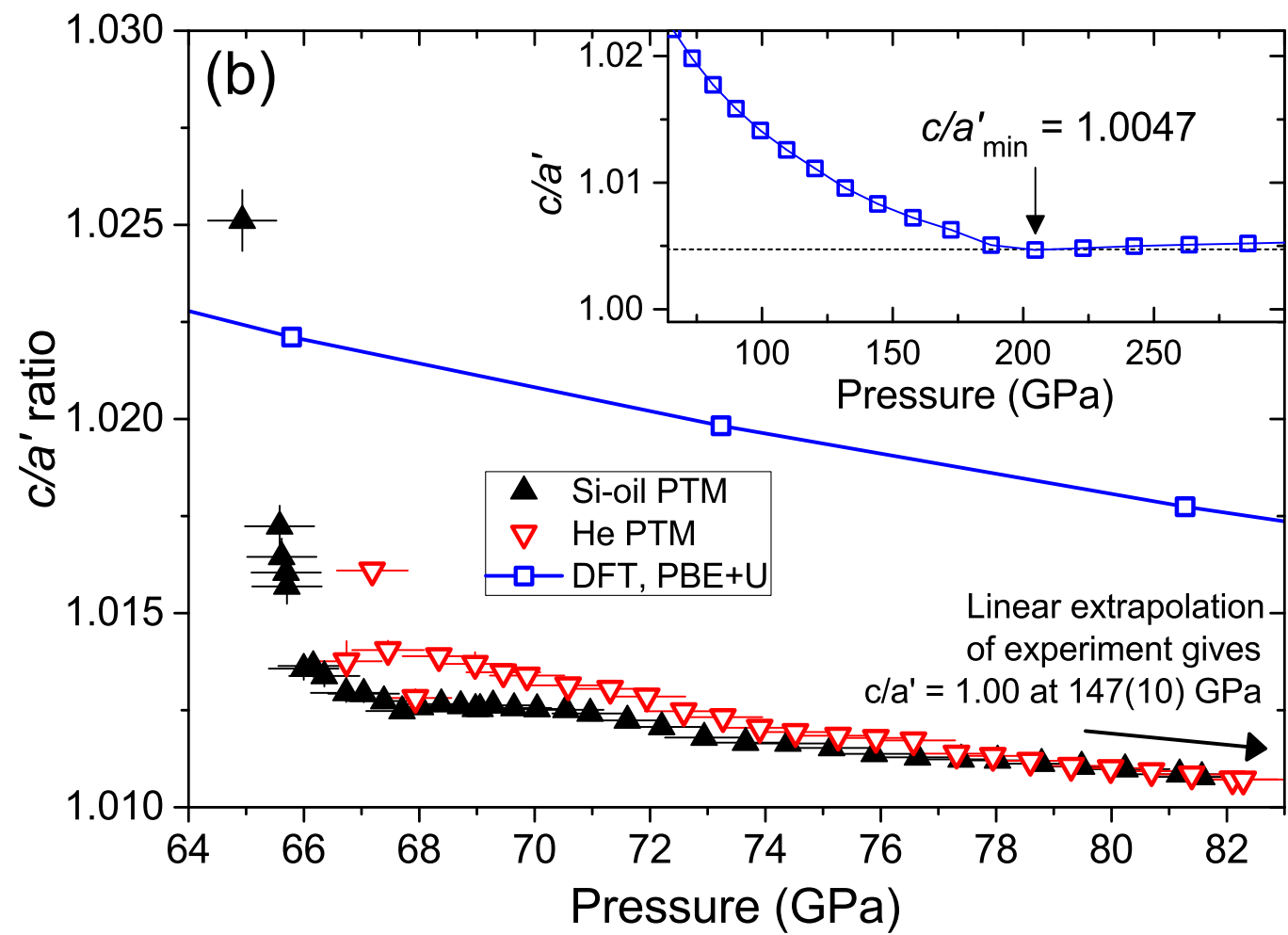

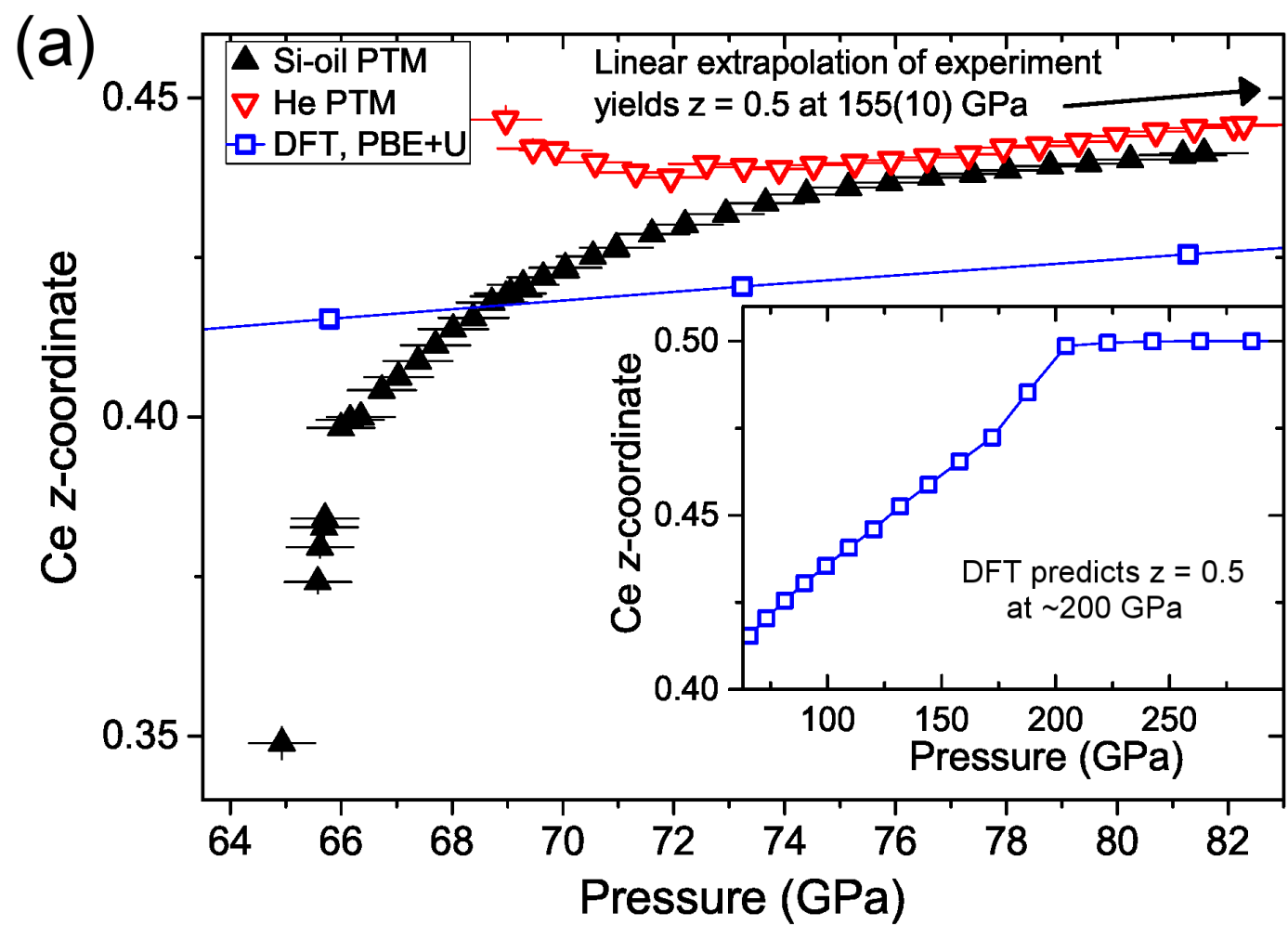 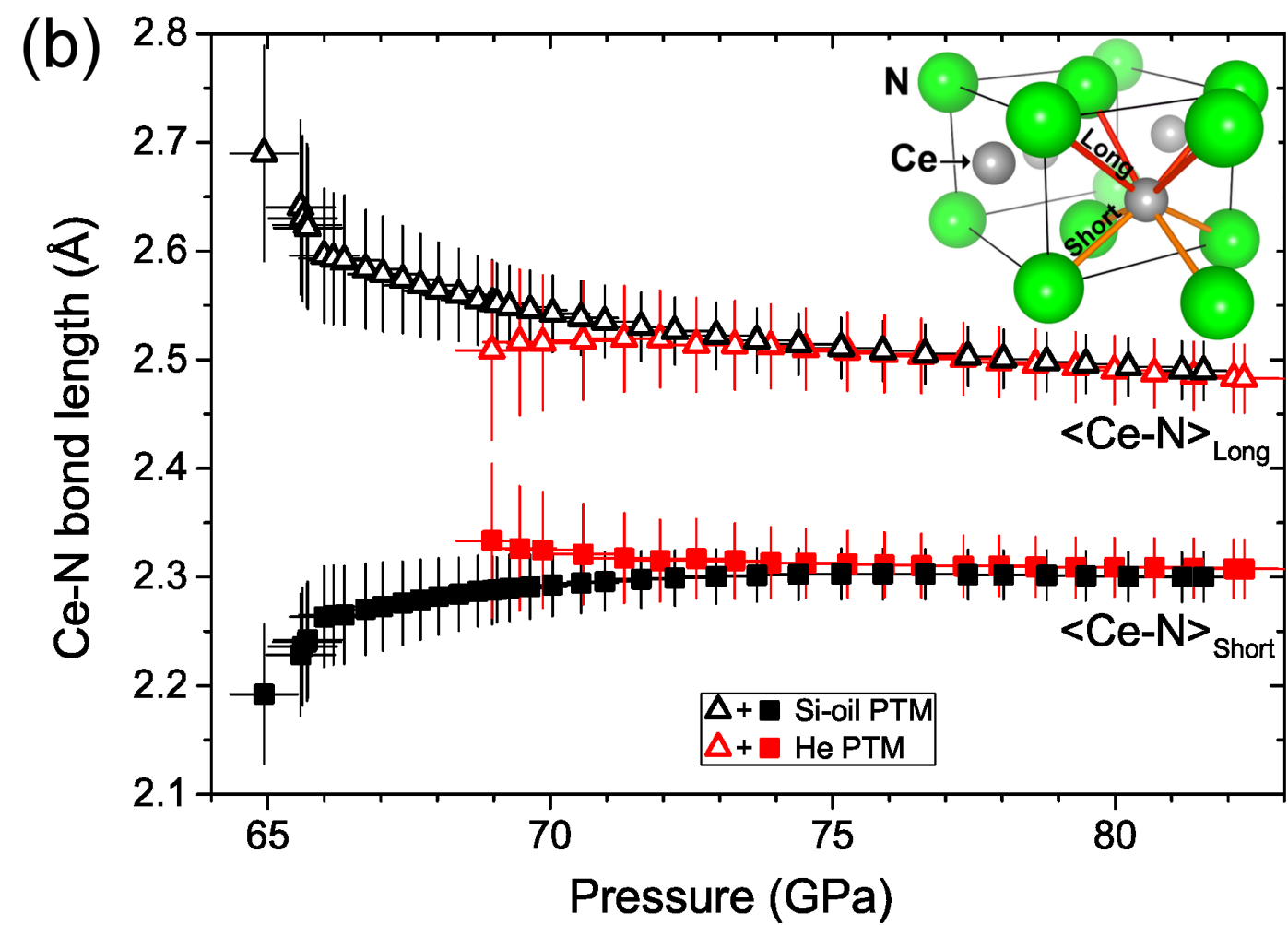

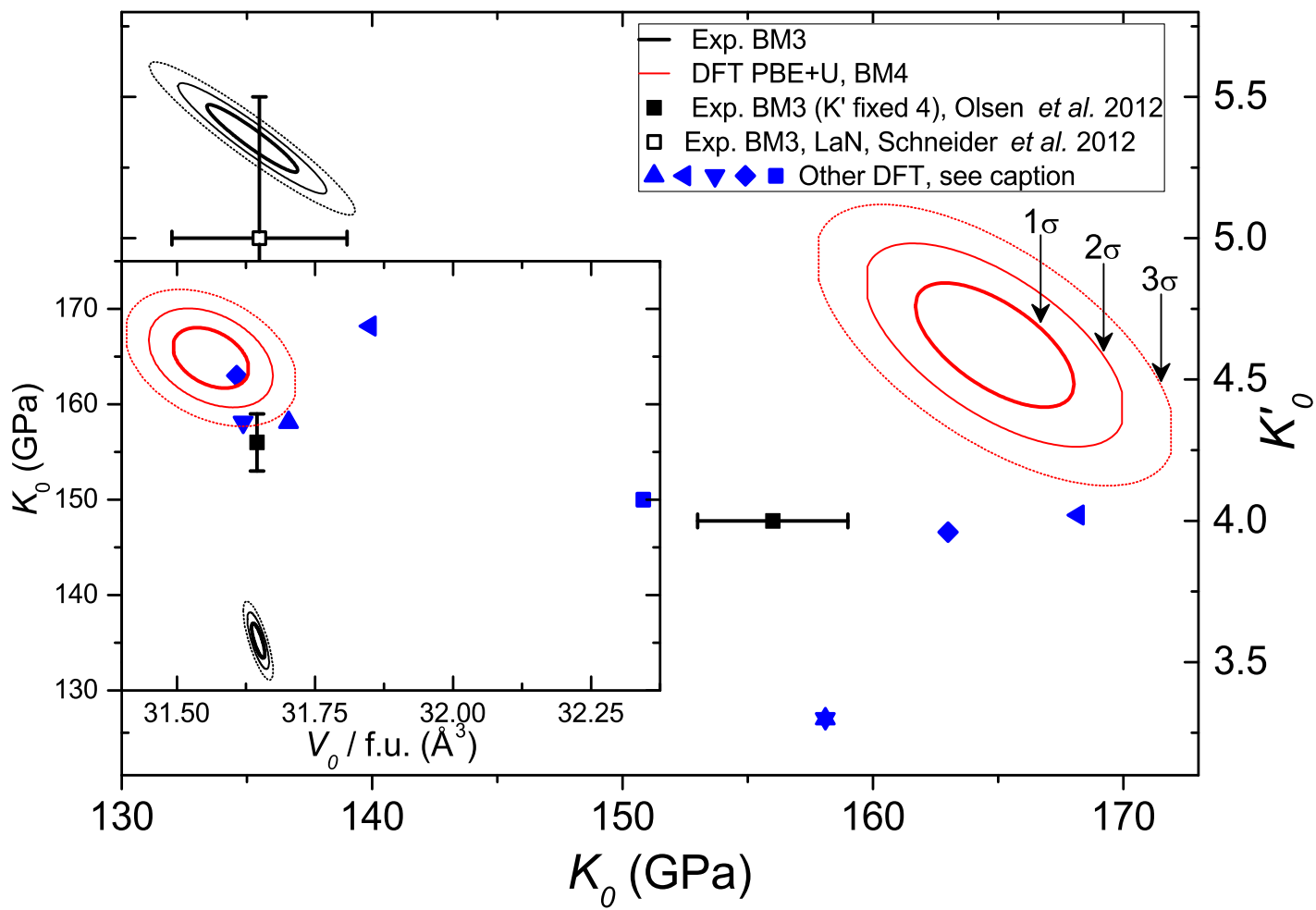

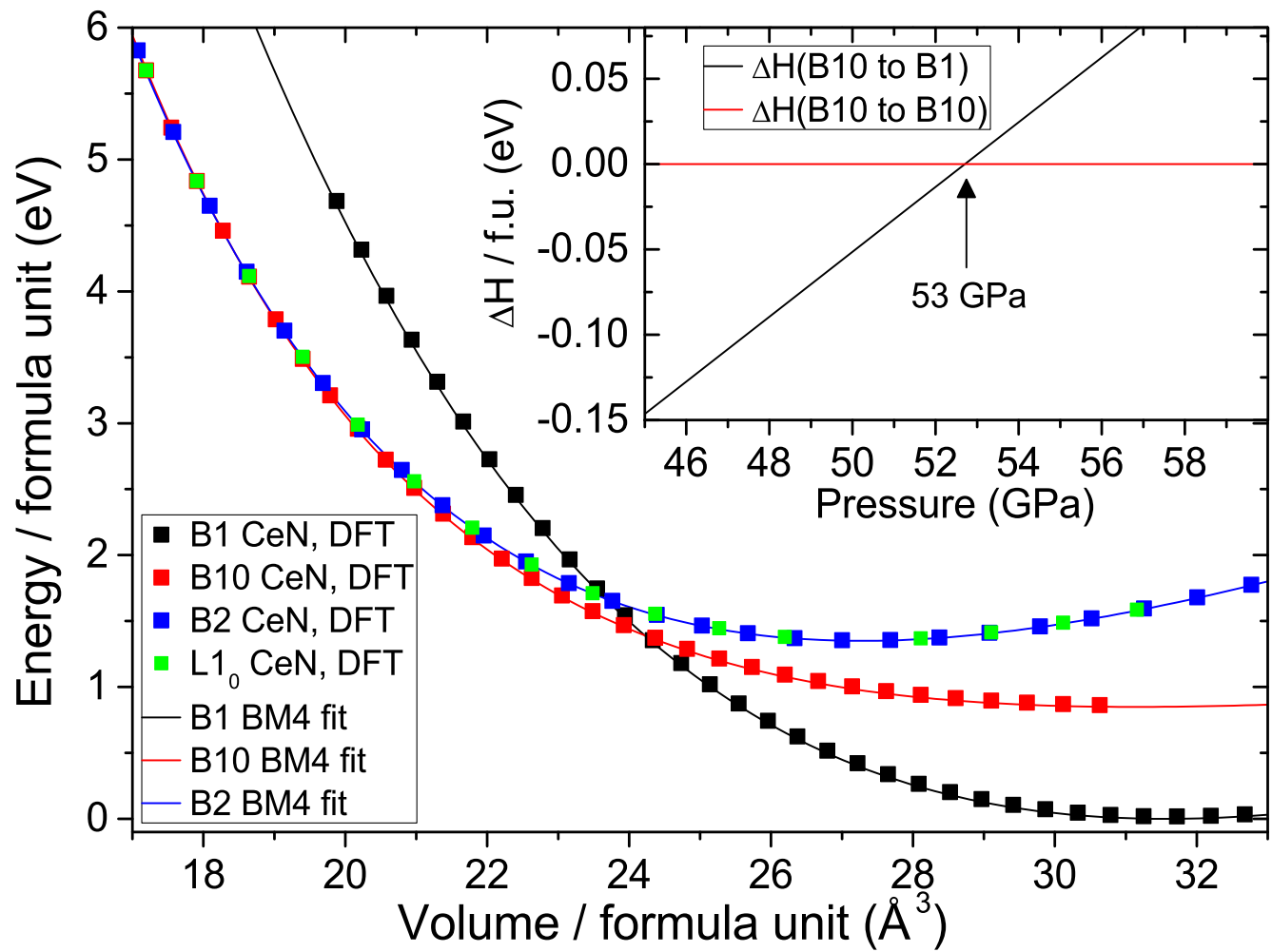